\documentclass[12pt,preprint]{aastex}
\shortauthors{Nisenson et al}

\begin{document}

\title{Motions of Isolated G-Band Bright Points in the Solar Photosphere}
\author{P. Nisenson and A. A. van Ballegooijen}
\affil{Harvard-Smithsonian Center for Astrophysics\\
60 Garden Street MS 15\\Cambridge, MA 02138}
\email{pnisenson@cfa.harvard.edu, vanballe@cfa.harvard.edu}
\and
\author{A. G. de Wijn and P. S\"{u}tterlin}
\affil{Sterrekundig Instituut\\ P.~O.~Box~80\,000\\
3508~TA~Utrecht\\ The~Netherlands}
\email{A.G.deWijn@astro.uu.nl, P.Suetterlin@astro.uu.nl}

\begin{abstract}
Magnetic elements on the quiet sun are buffeted by convective flows
that cause lateral motions on timescales of minutes. The magnetic
elements can be observed as bright points (BPs) in the G band at 4305
{\AA}. We present observations of BPs based on a long sequence of
G-band images recorded with the Dutch Open Telescope (DOT) and
post-processed using speckle masking techniques. From these images we
measured the proper motions of isolated BPs and derived the
auto-correlation function of their velocity relative to the solar
granulation pattern. The accuracy of BP position measurements is
estimated to be less than 23 km on the Sun. The rms velocity of BPs
(corrected for measurement errors) is about 0.89 km s$^{-1}$, and the
correlation time of BP motions is about 60 s.  This rms velocity is
about 3 times the velocity measured using cork tracking, almost
certainly due to the fact that isolated BPs move more rapidly than
clusters of BPs.  We also searched for evidence of vorticity in the
motions of G band BPs.
\end{abstract}


\section{Introduction}

Observations of the Sun with high spatial resolution show network
bright points (Muller 1983, 1985, 1994; Muller \& Keil 1983; Muller \&
Roudier 1984, 1992) and ``filigree'' (Dunn \& Zirker 1973; Mehltretter
1974; Berger et al 1995), which are chains of bright features located
within the intergranular lanes. The bright points and filigree are
seen in the wings of strong spectral lines such as H$\alpha$ and
Ca II H \& K, in lines formed in the photosphere, and even at
continuum wavelengths (with reduced contrast). The widths of these
structures is 100 to 200 km, at the limit of resolution of
ground-based solar telescopes. In the following we collectively refer
to these bright structures as bright points (BPs). The BPs are
associated with regions of strong magnetic field (Chapman \& Sheeley
1968; Title, Tarbell \& Topka 1987; Simon et al 1988;
Title et al 1989, 1992; Keller 1992) and correspond to magnetic flux
tubes of kilogauss field strength that stand nearly vertically in the
solar atmosphere (Stenflo 1973; Stenflo \& Harvey 1985; S\'{a}nchez
Almeida \& Mart\'{i}nez Pillet 1994; see review by Solanki 1993).

The dynamical behavior of BPs has been studied by a number of
authors. Muller (1983) found that facular points on the quiet sun
are predominantly located in patches at the periphery of supergranule
cells, indicating that the magnetic elements are advected by the
supergranular flow. The BPs always first appear in the dark
spaces at the junction of several granules, never inside a granule nor
in the space between only two granules. As the granulation pattern
evolves, the BPs remain in the intergranular spaces
throughout their lifetime, but not necessarily at the junction of
several granules like at the time of their first appearance. New
BPs have a tendency to appear adjacent to existing points,
and 15 \% of the BPs seem to split into two points which
move apart until a separation of 1 to {1.5\arcsec} is reached.  

Muller et al (1994) measured velocities of 29 isolated BPs and found
a mean speed of 1.33 km s$^{-1}$. Strous (1994) studied BPs in a growing
active region. Using line-center images taken in Fe~I (5576~{\AA}), he
found velocities between 0.26 km s$^{-1}$ and 0.62 km s$^{-1}$. Berger \& Title
(1996) measured velocities of 1 to 5 km s$^{-1}$ for G-band (4305~{\AA}) BPs
in the ``moat'' around a sunspot; they showed that the motions are
constrained to the intergranular lanes and are primarily driven by the
evolution of the granulation pattern. They found that the BPs
continually split up and merge, with a mean time between splitting
events of few hundred seconds. L\"{o}fdahl et al (1998) and Berger
et al (1998) analyzed G-band and continuum images obtained at the
Swedish Vacuum Solar Telescope (SVST) on La Palma and found rapid
splitting and merging of BPs in an enhanced network region.
Van Ballegooijen et al (1998) used a ``cork tracking'' method to
measure the proper motions of G-band BPs and found that BPs appear to
be passively advected by the granulation flow. The filigree are known
to be associated with abnormal granulation patterns (Dunn \& Zirker
1973), and the granules near network BPs are smaller and more numerous
than near a normal intergranular space (Muller, Hulot \& Roudier
1989), suggesting that the magnetic field has some effect on the
granulation flow. 

Little is known about the small-scale dynamics of flows in
intergranular lanes and the interaction of magnetic elements with
such flows. Theoretical models (e.g.,~Stein \& Nordlund 2000;
Emonet \& Cattaneo 2001) predict that vorticity is concentrated
in the lanes and that magnetic elements exhibit rapid rotational
motions. At present, there is no direct observational evidence for
vorticity in intergranular lanes. More detailed measurements of
vorticity are needed for testing magneto-convection models, and as
input for models of flux-tube waves that heat the upper solar
atmosphere (Hasan et al 2002).

In this paper we present measurements of BP proper motions within the
intergranular lanes. In particular, we look for evidence of vortical
motions. Tracking BPs is a difficult task. Long sequences of images
with very good seeing (or corrections of seeing) are needed. Image
jitter due to residual atmospheric effects needs to be reduced as much
as possible. The measurements of BP positions need to be made with
respect to some frame of reference.  The solar granulation is used as
a reference, and the motion of this reference frame is determined by
correlation tracking on the granulation pattern. We also tested
measuring the BPs with respect to the average position of the bright
points in each field, but found that the result is nearly identical to
the result from correlation tracking.  BPs vary widely in lifetime and
contrast, so it is very difficult to automate the process of tracking
them. In this paper, we use manual selection of BPs followed by
fitting the peaks.

\section{Observations}

Image sequences recorded with good seeing in the G band (4305~{\AA})
are essential for tracking BPs over many frames. Here we use G-band
images recorded with the Dutch Open Telescope (DOT) (Hammerschlag \&
Bettonvil 1998; Rutten et al 2001, 2002; S\"{u}tterlin et al
2001). The image sequence was collected on 2001 October 19 starting at
10:39 UT under good seeing conditions from a Network region close to
disk center. The pixel size for these images was {0.071\arcsec}
(corresponding to 51.5 km on the Sun), and the filter bandpass was
10~{\AA}.  The data were recorded at a rate of 6 frames per second,
and a burst of 100 frames was collected every 30 s. Each burst is
combined into a single reconstructed frame using speckle image
processing. The speckle processing uses an improved version of the
code of de Boer, Kneer \& Nesis (1992), who first applied the speckle
masking method of Weigelt (1977) to solar images, making use of the
spectral line ratio technique (von der L\"{u}he 1984). The resulting
199 frame sequence (1 frame every 30 seconds) achieves {0.2\arcsec}
resolution over the entire sequence. This is very close to the
diffraction limit of the 43.8 cm aperture. Figure \ref{f1a} is a
typical frame from the sequence.

The images in the sequence were crudely co-aligned by visual tracking
of a group of BPs. We then selected two subregions, each
256$\times$256 pixels in size, for measurement of BP positions (see
Figure \ref{f1a}). For each subregion we used correlation tracking to
determine a series of displacements of the granulation pattern in
consecutive frames. These displacements were measured with sub-pixel
accuracy by finding the peak of the cross-correlation of consecutive
subimages and fitting a quadratic function to the values near the
peak. We accumulated these displacements to obtain a series of
positions of the granulation pattern, $(x_{{\rm gran},n},
y_{{\rm gran},n})$, where $n$ is the frame index. We then went through
each frame many times, using a cursor to select a BP and follow it for
as many frames as we could, before it disappeared or merged with other
BPs. Each BP position was again measured with sub-pixel accuracy by
quadratic fitting, and the granulation position was subtracted. The
result was a list of BP positions $x_{k,n}$ and $y_{k,n}$ in a
reference frame tied to the solar granulation pattern, where $k$ is
the BP index. The advantage of this method is that an accurate
co-alignment and interpolation of the subimages is not required.

We only selected BPs that were clearly separated from other BPs so
that we could follow them without confusing their identity. We stopped
tracking them if they started to merge with other BPs. Any BPs that we
could not track for at least 3.5 minute (7 frames) were discarded. Our
measurements yielded 2992 positions for 161 BPs. The longest-living BP
we were able to track lived for 25 minutes (50 frames). The mean
lifetime was 9.2 minutes.

\begin{figure}[t]
\epsscale{0.8}
\plotone{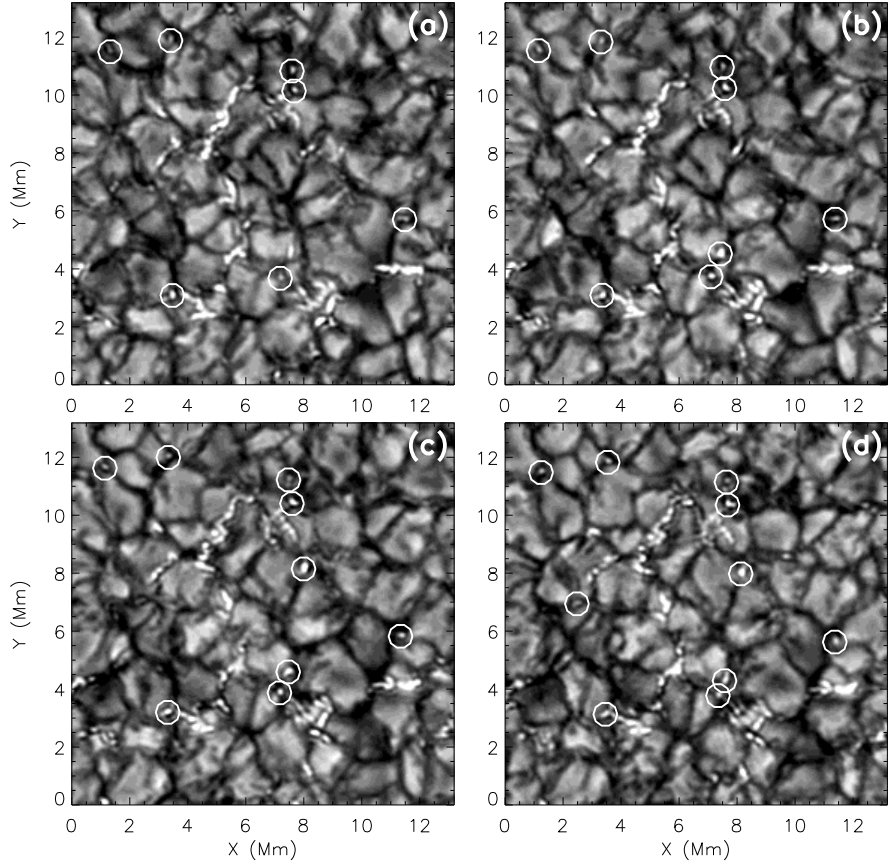}
\caption{Four images and with BP locations circled. The frames are
separated in time by 1 minute.}
\label{f2}
\end{figure}

\begin{figure}[t]
\epsscale{0.8}
\plotone{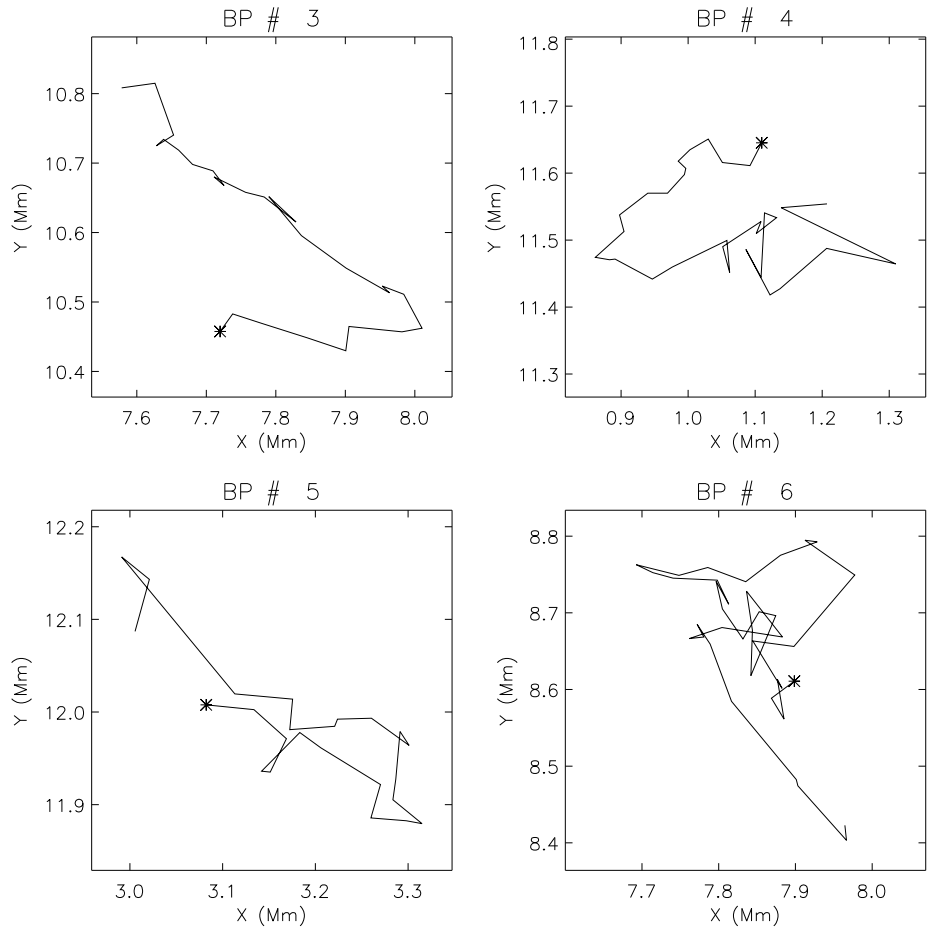}
\caption{Four bright point tracks. The starting point of each track
is indicated by a star.}
\label{f3}
\end{figure}

Figure \ref{f2} shows four subregion frames, separated in time by
1 minute. The circles show the positions of the BPs selected from each
frame. Figure \ref{f3} shows the paths of four individual BPs. In
some cases, like the upper-left and upper-right plots, the BP followed
a simple curved path. In other cases, the motions were erratic and
complicated, as in the lower-right plot. Some BPs move in nearly
straight lines, following intergranular lanes.

\section{Analysis}

\begin{figure}[t]
\epsscale{0.9}
\plotone{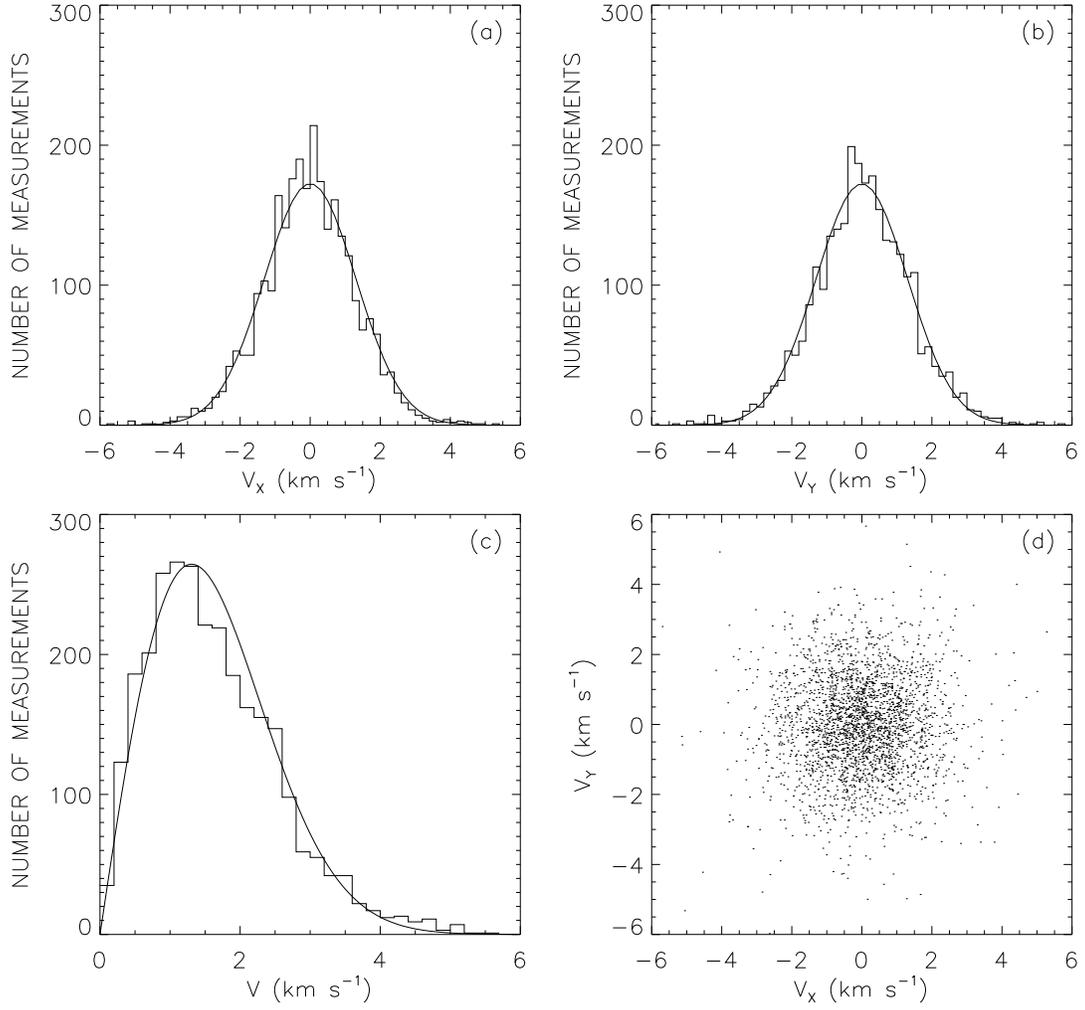}
\caption{Histograms of measured BP velocities: (a) $v_x$;
(b) $v_y$; (c) $v$. The results are compared with a Gaussian
distribution with rms velocity 1.31 km s$^{-1}$. Panel (d) shows
$v_x$ plotted against $v_y$.}
\label{f4}
\end{figure}

\begin{figure}[t]
\epsscale{0.9}
\plotone{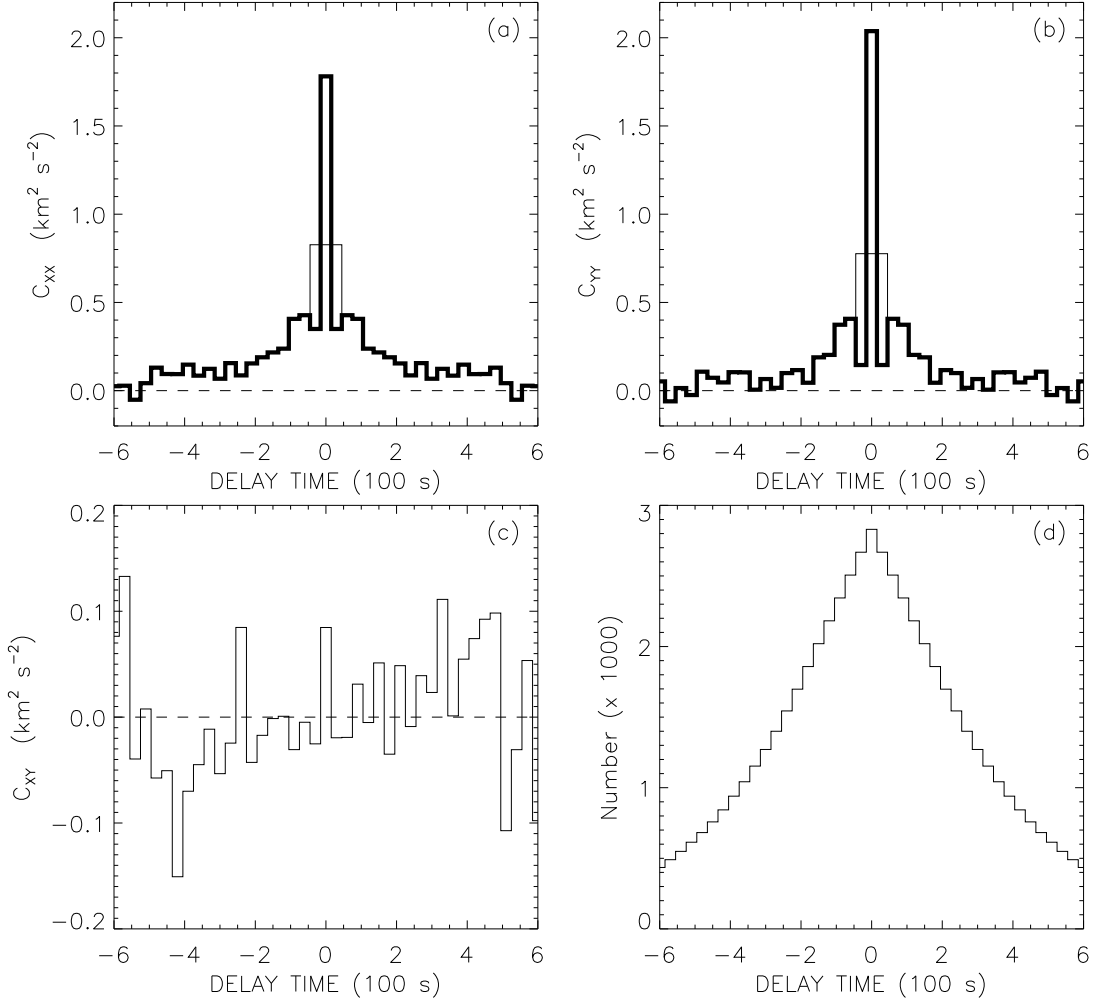}
\caption{Correlation functions of BP velocity $v_x$ and $v_y$.
(a) Observed auto-correlation $C_{xx,m}$ as function of delay time
$t = 30 m$ [s] ({\it thick curve}), and the average auto-correlation
$\tilde{C}_{xx}$ ({\it thin curve}). (b) Similar for
the observed auto-correlation $C_{yy,m}$ ({\it thick curve}) and
average correlation $\tilde{C}_{yy}$ ({\it thin curve}).
(c) Cross-correlation $C_{xy,m}$ as function of delay time $t$.
(d) Number of measurements per bin used in panels (a), (b) and (c).}
\label{f5}
\end{figure}

We calculated BP velocities $v_{x,k,n}$ and $v_{y,k,n}$ from position
differences over time intervals of 30 seconds. Figure \ref{f4} shows
histograms of $v_x$, $v_y$ and $v$ ($= \sqrt{v_x^2+v_y^2}$), and
a scatter plot of $v_x$ versus $v_y$. The histograms are consistent
with a Gaussian distribution with rms velocity of 1.31 km s$^{-1}$. This
includes both real signal and measurement errors. We also computed the
auto-correlations $C_{xx,m}$ and $C_{yy,m}$, and the cross-correlation
$C_{xy,m}$:
\begin{equation}
C_{xx,m} = \langle v_{x,k,n} v_{x,k,n+m} \rangle , ~~~~~
C_{yy,m} = \langle v_{y,k,n} v_{y,k,n+m} \rangle , ~~~~~
C_{xy,m} = \langle v_{x,k,n} v_{y,k,n+m} \rangle ,
\end{equation}
where $m$ is the number of time steps separating two velocity
measurements for the same BP, and the brackets denote an average over
the BP index $k$ and frame number $n$. Figures \ref{f5}a and \ref{f5}b
show the auto-correlations as function of the delay time $t = 30 m$
[s]. Figure \ref{f5}c shows the cross-correlation function. The
velocities $v_x$ and $v_y$ appear to be uncorrelated to within the
measurement errors (note the expanded vertical scale of this figure
compared to Figures \ref{f5}a and \ref{f5}b). Figure \ref{f5}d shows
the number of measurements used to derive these correlation functions.

Each auto-correlation function shows a strong peak at zero time lag,
and dips at lags of $\pm 1$ frame (30 seconds time delay). We suggest
that these features are due to the measurement errors. Let $u_{x,k,n}$
be the $x$-component of the true velocity of BP $k$ in the time
interval between frames $n$ and $n+1$, then the measured velocity is
\begin{equation}
v_{x,k,n} = u_{x,k,n} + \frac{\delta x_{k,n+1} - \delta x_{k,n}}
{\Delta t} ,
\end{equation}
where $\Delta t$ = 30 [s] is the time between frames, and $\delta
x_{k,n}$ are the errors in {\it position} measurement. Now assume that
these errors are randomly distributed with standard deviation $\sigma$
(the same for all BPs) and are uncorreleated from frame to frame. Then
the observed auto-correlation $C_{xx,m}$ is:
\begin{eqnarray}
C_{xx,0} & = & \langle u_{x,k,n}^2 \rangle + \sigma_{\rm v}^2 , 
\nonumber \\
C_{xx,\pm 1} & = & \langle u_{x,k,n} u_{x,k,n \pm 1} \rangle
- \sigma_{\rm v}^2 /2 , \\
C_{xx,m} & = & \langle u_{x,k,n} u_{x,k,n+m} \rangle
~~~~~~~ \hbox{for} ~~ | m | \geq 2 , \nonumber
\end{eqnarray}
and similar for $C_{yy,m}$,
where $\sigma_{\rm v} = \sigma \sqrt{2} / \Delta t$ is the error
in the velocity measurement. The position errors increase the observed
correlation at $m=0$ by $\sigma_{\rm v}^2$ and reduce the correlation
at $m = \pm 1$ by $- \sigma_{\rm v}^2 /2$. Therefore, the {\it
average} auto-correlations, defined by
\begin{equation}
\tilde{C}_{xx} \equiv \frac{1}{3} \sum_{m=-1}^{+1} C_{xx,m} , ~~~~~
\tilde{C}_{yy} \equiv \frac{1}{3} \sum_{m=-1}^{+1} C_{yy,m} ,
\end{equation}
are not systematically affected by the measurement errors. These
average correlations are indicated by the thin lines in
Figures \ref{f5}a and \ref{f5}b. The difference between $C_{xx,0}$ and
$\tilde{C}_{xx}$ provides an upper limit on $\sigma_{\rm v}^2$ because
the true auto-correlation ${\langle u_{x,k,n} u_{x,k,n+m} \rangle}$
decreases with increasing $|m|$. The auto-correlations for $v_x$ and
$v_y$ yield upper limits on $\sigma_{\rm v}$ of 0.98 km s$^{-1}$ and 1.12
km s$^{-1}$, respectively. Therefore, the upper limit on the position error
$\sigma$ is about 24 km.

Figures \ref{f5}a and \ref{f5}b show that the auto-correlations are
significantly different from zero for delays up to 500 s. The average
correlations for are $\tilde{C}_{xx} = {0.83 \pm
0.03}$ $\rm km^2 s^{-2}$ and $\tilde{C}_{yy} = {0.78 \pm 0.03}$
$\rm km^2 s^{-2}$, where the error estimates are based on the
cross-correlations shown in Figure \ref{f5}c. This provides a lower
limit on the true velocity dispersion $u_0$ of BPs in one direction:
$u_0 \approx 0.89$ km s$^{-1}$ based on the average of $\tilde{C}_{xx}$ and
$\tilde{C}_{yy}$. Furthermore, $\tilde{C}_{xx}$ and $\tilde{C}_{yy}$
are about twice the auto-correlations at $m = \pm 2$. Therefore,
the correlation time of the BP velocity is about 60 seconds
(2 frames).

\begin{figure}[t]
\epsscale{0.9}
\plotone{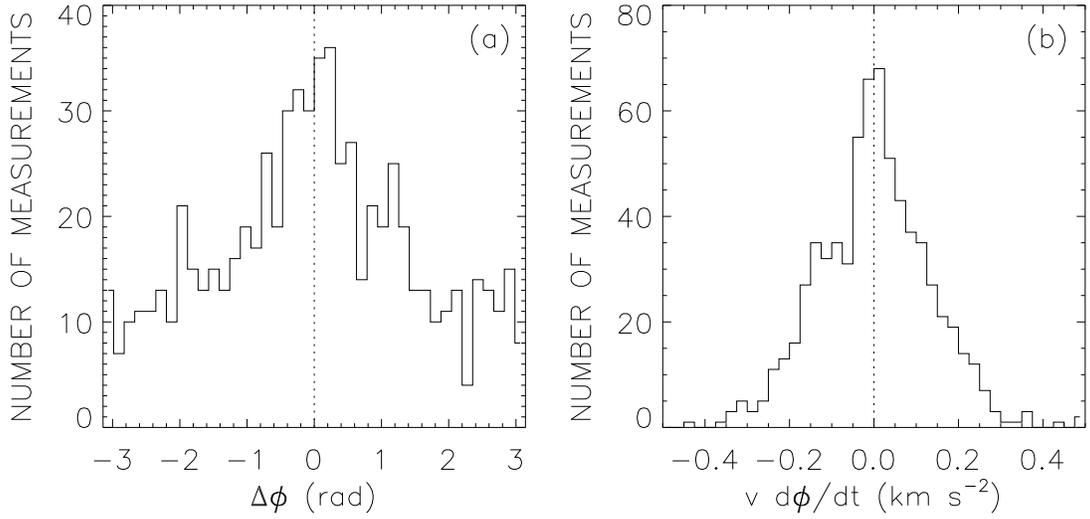}
\caption{(a) Histogram of the change $\Delta \phi$ in direction angle
of BP velocity. (b) Histogram of the centrifugal acceleration
$v d \phi /dt$.}
\label{f6}
\end{figure}

We were interested in using BPs as a possible diagnostic of vortical
motions in the intergranular lanes. A BP caught up in a vortex within
an intergranular lane might exhibit rotational motions. In an effort
to detect such motions, we computed changes in direction angle,
$\Delta \phi$, of BP velocity in two neighboring time intervals.  For
this calculation we discarded any measurements for which $v < 1.5$
km s$^{-1}$ in either time interval, since the angle errors became large for
smaller velocities.  This left us with 687 measurements. A histogram
of $\Delta \phi$ is shown in Figure \ref{f6}a. The distribution shows
a broad peak at $\Delta \phi = 0$, indicating that the direction of
motion persists from one time interval to the next. Figure \ref{f6}b
shows a histogram of the centrifugal acceleration, $v d\phi /dt$,
which is the relevant quantity when considering the generation of
transverse waves in magnetic flux tubes. Histograms such as these
could be useful tools for comparing models of flux tube waves to
observations.

The above measurements use the granulation pattern as a reference
frame. In order to check whether the granulation provides a stable
reference frame, we also measured BP positions relative to each other
(independent of the granulation pattern). For this purpose we
computed a reference position $(x_{{\rm ref},n}, y_{{\rm ref},n})$
as follows. For each pair of frames $n$ and $n+1$, we determine the
average positions $(\bar{x}_n,\bar{y}_n)$ and $(\bar{x}_{n+1},
\bar{y}_{n+1})$ of all BPs {\it in common} between these consecutive
frames. Then we accumulate the position differences:
\begin{equation}
x_{{\rm ref},n+1} = x_{{\rm ref},n} + \bar{x}_{n+1} - \bar{x}_n ,
~~~~~
y_{{\rm ref},n+1} = y_{{\rm ref},n} + \bar{y}_{n+1} - \bar{y}_n ,
\end{equation}
starting with $x_{{\rm ref},0} = y_{{\rm ref},0} = 0$ in the first
frame. This provides a reference position against which the individual
BPs can be measured. The relative positions are $x_{k,n}^\prime
= x_{k,n} - x_{{\rm ref},n}$ and $y_{k,n}^\prime = y_{k,n} -
y_{{\rm ref},n}$. The auto-correlations of $v_x^\prime$ and
$v_y^\prime$ yield $\tilde{C}_{xx} = {0.71 \pm 0.03}$ $\rm km^2
s^{-2}$ and $\tilde{C}_{yy} = {0.66 \pm 0.03}$ $\rm km^2 s^{-2}$,
which is about 15\% less than the values derived for the granulation
reference frame. This reduction in BP velocities is to be expected
because there are on average only 7.0 BPs in common between
consecutive frames, and subtracting the mean velocity of these 7 BPs
should reduce $\tilde{C}_{xx}$ and $\tilde{C}_{yy}$ by a factor $6/7$.
The fact that the observed reduction is not any larger than predicted
implies that there is nothing to be gained by measuring the BP motions
relative to each other, and in fact this causes the BP velocities
to be systematically underestimated. We conclude that the granulation
pattern provides a stable reference frame for measurement of BP
motions.

\section{Discussion}

We have shown that with good enough data we can precisely measure
positions for isolated BPs, obtaining both their linear and angular
motions. The errors due to image jitter have been reduced to levels
of 25 km (0.5~pixels) or less. This allows us to measure the motions
of individual BPs to a precision of less than 1 km s$^{-1}$ for each 30
second time interval, which will be important for studying the
generation of MHD waves in flux tubes. We also track BPs for
sufficiently long times that we can see the regularity in their
motions.

The rms velocities we measured for isolated BPs were almost a factor
of three higher than velocities measured using cork tracking (van
Ballegooijen et al 1998). We believe this is due to a selection
effect. By only measuring isolated BPs we measure the fastest-moving
features. Once BPs cluster with other BPs, mostly at lane interstices,
they seem to be ``captured'' by the group and their motions are
reduced. There are many more BPs in these clusters than there are
isolated BPs so the cluster statistics dominate when measurements are
made with cork tracking and the average velocity is reduced.

In measuring the motions of BPs, we have found that they move in
somewhat circular paths. Combining the angular change of their motions
with the distance they travel is a potential way of estimating the
(vertical) vorticity in the flow field, assuming the BPs act as test
particles. While performing the tracking of individual BPs, we saw
little evidence of vorticity other than the tracks of the BPs
themselves.  We looked for, but did not find, pairs of BPs that
orbited one another; in fact, BPs that came in close proximity to
other BPs tended to have reduced motion. We also saw little evidence
of swirling motions in the granules or granular lanes that would
correspond to the observed BP motions. This might indicate that BPs
are affected not only by surface flows associated with the solar
granulation, but also by other flows that occur at larger depth below
the photosphere.  Further modeling will be required to establish
whether or not the BPs move passively with the surface flow, and to
determine whether the observed BP motions are consistent with models
of flux tube waves and heating of the upper solar atmosphere.

\acknowledgements

The authors wish to thank R. J. Rutten, R. Hammerschlag and
F. Bettonvil for the use of DOT solar data. We also wish to especially
thank R.J. Rutten for his very helpful suggestions on an earlier draft
of this paper.  We wish to acknowledge support for the work from the
National Science Foundation (Grant ATM-9811523) to the Smithsonian
Astrophysical Observatory.

\newpage

\begin{figure}[ht]
\epsscale{0.9}
\plotone{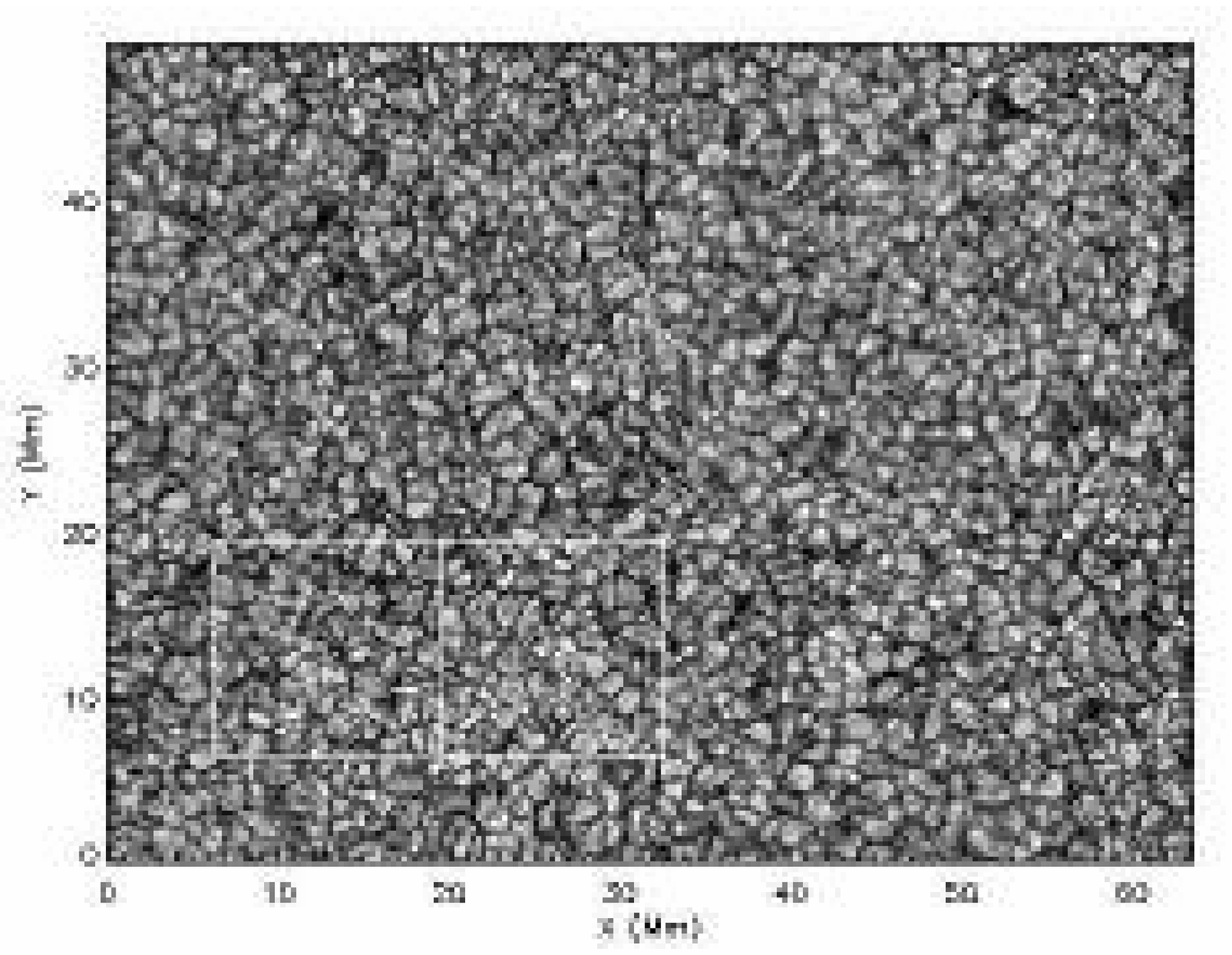}
\caption{G-band frame taken at 10:39:35 UT. The boxes show the two
subregions used for measurement of BP positions.}
\label{f1}
\end{figure}

\end{document}